**Electrostatic treatment of charged interfaces in classical atomistic simulations**

Cong Tao[1,2], Daniel Mutter[1], Daniel F. Urban[1], and Christian Elsässer[1,3]

[1]Fraunhofer IWM, Wöhlerstraße 11, 79108 Freiburg, Germany

[2]Institute of Applied Materials-Computational Materials Science (IAM-CMS), Karlsruhe Institute of Technology, Straße am Forum 7, 76131 Karlsruhe, Germany

[3]Freiburg Materials Research Center (FMF), University of Freiburg, Stefan-Meier-Straße 21, 79104 Freiburg, Germany

**Abstract**

Artificial electrostatic potentials can be present in supercells constructed for atomistic simulations of surfaces and interfaces in ionic crystals. Treating the ions as point charges, we systematically derive an electrostatic formalism for model systems of increasing complexity, both neutral and charged, and with either open or periodic boundary conditions. This allows to correctly interpret results of classical atomistic simulations which are directly affected by the appearance of these potentials. We demonstrate our approach at the example of a strontium titanite ($SrTiO_3$) supercell containing an asymmetric tilt grain boundary. The formation energies of charged oxygen vacancies and the relaxed interface structure are calculated based on an interatomic rigid-ion potential, and the results are analyzed in consideration of the electrostatic effects.

# 1  Introduction

Electrostatic effects play a significant role in atomistic simulations of ionic and partly ionic crystals containing surfaces [1–3], grain boundaries [4–6], or interfaces between different materials [7–9]. A supercell for the simulation of such a system is composed of planes, which can be electrically neutral or charged, depending on their orientation and elemental occupation. A fundamental classification of the stacking of charged planes of overall neutral ionic crystals was given by Tasker [10], who distinguished three generic types of configurations: in the first type, the planes themselves are neutral, consisting of anions and cations in stoichiometric composition [as in the rock salt structure in (100) orientation]. In the second type, the planes are charged, but can be grouped into neutral, mirror symmetric arrangements (repeat units) of three planes, e.g., with the charge sequence $-q, +2q, -q$ [as in the fluorite structure in (111) orientation]. In the third type, the planes are alternatingly charged, i.e., the neutral repeat units consist of two planes with the charge sequence $+q, -q$ [as in the zincblende structure in (100) orientation]. Moreover, more complex stacking sequences than those defined by Tasker can be present in crystals, consisting of more than two or three planes, as e.g. the sequence of (111) planes in magnetite ($Fe_3O_4$) [11,12].

If a charge neutral crystal is terminated by surfaces oriented as the stacked planes, a stacking of the third type of Tasker's classification would result in two oppositely charged surface planes (polar surfaces [13]). This corresponds to an electric dipole moment perpendicular to the surfaces, which would become infinitely large for macroscopic slab thicknesses. Note that the term "dipole" is used in this macroscopic sense throughout the manuscript, not to be confused with the "microscopic", internal dipole moments between the neighboring alternatingly charged planes in the bulk of the crystal.

In the context of heterojunctions between polar systems (i.e., systems consisting of alternatingly charged planes) and non-polar systems (neutral planes), the scenario of a diverging electrostatic energy with the slab thickness is referred to as the "polar catastrophe" [14,15]. It leads to compensating effects which stabilize the interface, such as a change of the stoichiometry [7,16], e.g., by the formation of vacancies or adatoms, or an electronic redistribution [17–19]. Note that just a structural reconstruction by maintaining the stoichiometry cannot lead to a vanishing dipole moment.

A change of the stoichiometry at surfaces or interfaces leads to configurations which cannot be any longer completely described by the Tasker type three scenario. Such interfaces and the correct treatment of the excess species were described in detail by Finnis [20], who introduced the concept of tapered termination planes to deal with the ambiguity of calculating reasonable surface energies from atomistic supercell models.

In contrast, in the present work, we consider Tasker type three systems with perfectly stoichiometric surfaces and interfaces, which, as described above, exhibit a dipole moment corresponding to a non-vanishing electric field within the supercell. In addition, there is an electric field in the cell if it carries a total net charge and is treated under periodic boundary conditions in the simulation model. All these electric fields are artificial, since they wouldn't be present in real crystals. Therefore, in the model, the effects of the fields must be compensated "externally". Otherwise, their interaction with charged species (ions) results in artificial effects, e.g., on the equilibrium structure or on the formation energies of ionic defects.

If the supercell is considered isolated in the sense that open boundary conditions are applied in the direction perpendicular to the surfaces, a constant internal electric field can simply be removed by "switching on" an artificial external field in the opposite direction, as described by Meyer and Vanderbilt [2]. However, periodic boundary conditions are generally applied in atomistic simulations, with supercells containing a vacuum of finite length ("slab systems"). In this case, there is an additional contribution to the electric field ensuring the periodicity of the corresponding potential [2,21,22]. This field would vanish for an infinitely large vacuum size, which is practically impossible in the supercell approach. A widely applied method to circumvent this problem is to place an artificial dipole layer in the vacuum region to compensate the unwanted field [23,24]. In the context of density functional theory (DFT) based calculations, more advanced methods were proposed recently, where the potential is adapted within the DFT formalism to correctly deal with surface charges in a slab and a possible total charge of the simulation cell [25–27].

These approaches cannot be directly applied to systems where the interaction is described by classical interatomic potentials and where point charges are typically considered instead of charge densities. Of course, such a treatment does not allow for the study of electrostatic dipoles appearing even at formally charge neutral interfaces due to an inhomogeneous electron distribution, as seen for, e.g., Si/Ge [28], Si/SrTiO$_3$ [29], Pd/SrTiO$_3$ [30] or TiO$_2$/ZnO interfaces [31] studied by DFT. However, with the classical approaches, one is able to model more complex structures with a considerably larger number of atoms, needed, e.g., to simulate low angle grain boundaries [32] or polycrystalline systems containing multiple interfaces and/or surfaces.

In a recent work [33], we derived the electrostatic potential in supercells of the perovskite oxide SrTiO$_3$ (STO), modelled by a classical interaction potential, where the symmetric tilt grain boundaries $\Sigma 5(210)[001]$ and $\Sigma 5(310)[001]$ were present. For this purpose, we adapted the surrogate model of Freysoldt and Neugebauer [34], which describes the electrostatic potential near an interface by a properly placed effective point charge, to a system containing grain boundaries and surfaces. In the present work, we directly solve the one-dimensional Poisson equation for a supercell composed of homogeneously charged planes, for both, open and periodic boundary conditions. This approach has the advantage to be straightforwardly adaptable to complex cells containing any combination interfaces and surfaces. In addition to neutral systems, charged cells can be considered by accounting for the influence of a neutralizing background, which is commonly applied in the case of periodic boundary conditions to deal with a diverging electrostatic energy.

A supercell of STO containing the asymmetric tilt grain boundary (ATGB) (430)||(100) serves as a demonstration example in this study. Grain boundaries of this type have been experimentally observed in polycrystalline STO microstructures [35–37]. They are of special interest since their interaction with oxygen vacancies under the influence of external electric fields is assumed to be the origin of field-assisted grain growth in STO [38,39]. When studying this system by classical atomistic simulations, it is crucial to correctly take the artificial, i.e., non-compensated electrostatic potentials in the supercell into account in the analysis of the results, since they affect the atomic structure and the formation energies of charged oxygen vacancies.

The paper is organized as follows: in Section 2, we derive the electrostatic potential in a general supercell consisting of homogeneously charged lattice planes based on the Poisson equation. In Section 3, the formalism is exemplarily applied to generic systems containing alternatingly charged lattice planes (Tasker type 3) of increasing complexity, namely a charge-neutral and a charged cell of a single crystal, and a charged cell with an interface between two different crystals or differently oriented parts of a crystal. In Section 4, we determine the electrostatic potential specifically in an ideal and in a structurally relaxed STO supercell containing an ATGB (430)||(100) by calculating the formation energy of oxygen vacancies. In addition, the influence of the potential on the structure is discussed. In Section 5, we give a summary and concluding remarks.

## 2 Derivation of the electrostatic formalism

To derive the electrostatic model in the most general way, we compose a three-dimensional atomistic supercell of a crystal, which may contain surfaces, grain boundaries, or other kinds of interfaces, by a sequence of – in general – charged lattice planes with arbitrary distances to each other, as schematically sketched in Figure 1.

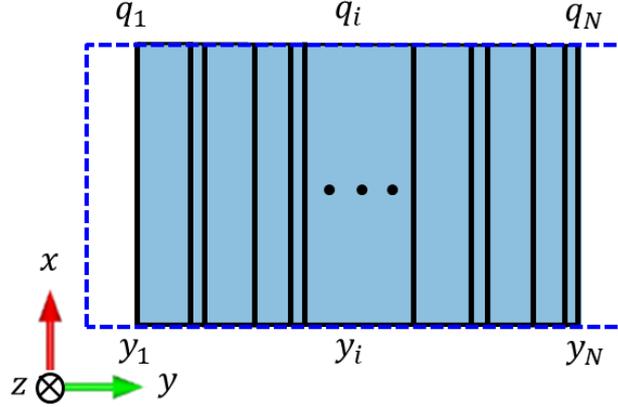

Figure 1. A two-dimensional sketch of a general atomistic supercell of a crystal (blue) containing charged ($q_i$) lattice planes at positions $y_i$. The dashed blue box represents the supercell, which may also contain vacuum.

In such a description, it is reasonable to apply periodic boundary conditions in the directions parallel to the lattice planes (here, the $x$ and $z$ directions), i.e., considering them as infinitely extended. Perpendicular to the lattice planes (here, the $y$ direction), either open (OBC) or periodic boundary conditions (PBC) can be applied. For example, to simulate surfaces of crystals, we can use slabs of atoms in supercells with OBC (single slabs with semi-infinitely extended vacuum regions outside their surfaces) or with PBC (periodically repeated slabs, separated by vacuum regions of finite thickness). The meaning of *open boundary conditions* in this regard is not to be confused with its meaning in the context of grand-canonical simulations, where the system is *open* to the exchange of atoms with external reservoirs based on the values of chemical potentials. In the simulations described in this work, the stoichiometry is always kept fixed. In the following, when referring to OBC or PBC, they are always meant to be applied in $y$ direction. In a one-dimensional continuum description, an $x$-$z$-plane averaged charge density $\rho(y)$ leads to an $x$-$z$-plane-averaged electrostatic potential $V(y)$ in the supercell, following the one-dimensional Poisson equation [40]:

$$\frac{d^2 V(y)}{dy^2} = -\frac{\rho(y)}{\epsilon}. \qquad (1)$$

We write $\rho(y)$ in the form $\rho(y) = \rho_0 + \rho_1(y)$, where $\rho_0$ denotes a homogeneous background charge density, such as the neutralizing (charge-compensating) background charge density in the case of a charged supercell, and $\rho_1(y)$ is the density profile resulting from the charged lattice planes. In a rigid ion model, where the ions are modelled as isolated point charges, a homogeneous charge density of a plane can be approximated as $q/A$, with $q$ being the total charge of the individual ions on an $x$-$z$-plane of the supercell with area $A$. The permittivity $\epsilon$ is equal to the vacuum permittivity $\epsilon_0$ in this model.

The charge density $\rho(y)$ of a system of $N$ charged planes located at positions $y_i$ can then be expressed as:

$$\rho(y) = \rho_0 + \sum_{i=1}^{N} \frac{q_i}{A} \cdot \delta(y - y_i), \qquad (2)$$

with the Dirac $\delta$-function. For this charge density, the general solution of the Poisson equation is given as:

$$V(y) = -\frac{\rho_0}{2\epsilon_0} y^2 - \sum_{i=1}^{N} \frac{q_i}{2A\epsilon_0} |y - y_i| + C_1 y + C_2, \qquad (3)$$

where $C_1$ and $C_2$ are constants of integration which are determined by the choice of boundary conditions. The negative derivative of $V(y)$ with respect to $y$ is the electric field $E(y)$:

$$E(y) = \frac{\rho_0}{\epsilon_0} y + \sum_{i=1}^{N} \frac{q_i}{2A\epsilon_0} \cdot \mathrm{sgn}(y - y_i) + C_1. \qquad (4)$$

Here, $\mathrm{sgn}(y)$ denotes the sign function, which can be expressed as a sum of two Heaviside step functions: $\mathrm{sgn}(y) = H(y) - H(-y)$. Differentiating Equation (4) with respect to $y$ and using $H'(y) = \delta(y)$ results in the density given in Equation (2) and thus Equation (1) is satisfied.

We first consider OBC. Without an external source of charge, there is no need to introduce a neutralizing background charge density ($\rho_0 = 0$), and without an additional external electric field (i.e., $C_1 = 0$), the potential reads

$$V_{\mathrm{OBC}}(y) = -\sum_{i=1}^{N} \frac{q_i}{2A\epsilon_0} |y - y_i| + C_2, \qquad (5)$$

where $C_2$ can be chosen arbitrarily.

Next, we consider PBC. Let the total charge of the supercell be $Q_{\mathrm{tot}} = \sum_{i=1}^{N} q_i$. For $Q_{\mathrm{tot}} \neq 0$, the infinite repetition of the supercell results in an undefined electrostatic energy. This problem is commonly treated by adding an averaged compensating background charge density $\rho_0 = -\frac{Q_{\mathrm{tot}}}{V_{\mathrm{cell}}}$ to the system. Here, $V_{\mathrm{cell}}$ is the volume of the supercell. To simplify the following derivations, we select the origin of the $y$ axis as the center of the supercell. PBC require that the potential is periodic, too, $V\left(-\frac{L}{2}\right) = V\left(+\frac{L}{2}\right)$, where $L$ denotes the length of the supercell in $y$ direction. From Equation (3), one obtains $C_1 = -\sum_{i=1}^{N} \frac{q_i y_i}{\epsilon_0 V_{\mathrm{cell}}}$, which effectively describes an internal electric field imposed by the PBC. In general, $C_1$ is non-zero, even for $Q_{\mathrm{tot}} = 0$.

The total potential is then given by:

$$V_{\mathrm{PBC}}(y) = V_{\mathrm{OBC}}(y) + \frac{Q_{\mathrm{tot}}}{2\epsilon_0 V_{\mathrm{cell}}} y^2 - \sum_{i=1}^{N} \frac{q_i y_i}{\epsilon_0 V_{\mathrm{cell}}} \cdot y + C. \qquad (6)$$

If $Q_{\text{tot}} \neq 0$, the coefficient of the linear term in $y$ can be written as $-\bar{y}Q_{\text{tot}}/\epsilon_0 V_{\text{cell}}$ with the "center of charge" given as $\bar{y} = \sum_{i=1}^{N} q_i y_i / \sum_{i=1}^{N} q_i$. Combining this with the quadratic term leads to an alternative formulation of $V_{\text{PBC}}(y)$:

$$V_{\text{PBC}}(y) = V_{\text{OBC}}(y) + \frac{Q_{\text{tot}}}{2\epsilon_0 V_{\text{cell}}}(y - \bar{y})^2 + C', \qquad (7)$$

with a shifted constant $C' = C - \frac{Q_{\text{tot}}}{2\epsilon_0 V_{\text{cell}}} \bar{y}^2$. In this way it is obvious that PBC in a non-neutral cell lead to an additional quadratic term in the electrostatic potential which is centered around the center of charge $\bar{y}$. Considering a single charged plane, the potential derived from Eq. (7) is consistent to the expression reported by Rutter (Eq. (1) in Ref. [27]).

## 3    The electrostatic potential for generic scenarios

A.  <u>Charge neutral supercell with equidistant, alternatingly charged planes</u>

We first consider the electrostatic potential for a cell of the type sketched in Figure 1. Specifically, the cell consists of an even number of $N$ equidistantly spaced (distance $d$) planes with alternating plane-averaged charge densities $\rho_i = (-1)^i |q|/A$ at the positions $y_i$ ($i = 1, \ldots, N$). Such a structure is apparently charge neutral ($Q_{\text{tot}} = 0$). This configuration, classified by Tasker as type 3, is realized e.g. in slab supercells for the atomistic treatment of (001) surfaces of the perovskites $BaTiO_3$ and $PbTiO_3$ [2], of (0001) surfaces of ZnO [18,41,42], or of (001) surfaces of $CeO_2$ [43,44].

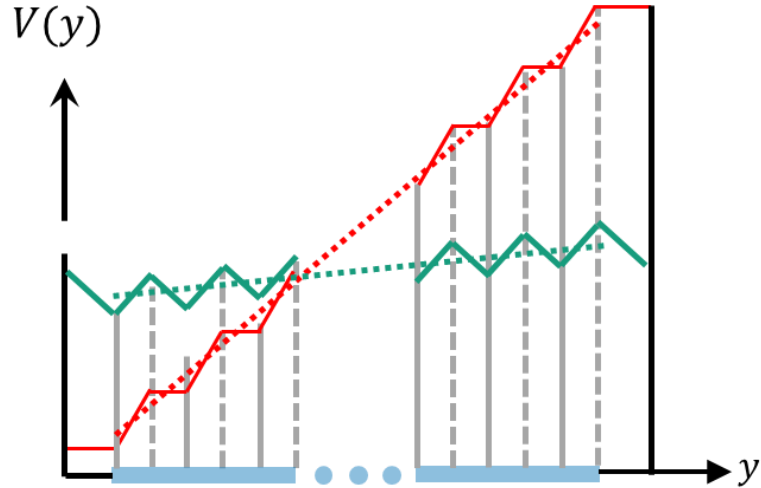

Figure 2. The electrostatic potential within a neutral cell with equidistant, alternatingly charged planes [indicated by solid (negative) and dashed (positive) vertical gray lines] for OBC (red line) and PBC (green line). The macroscopic average potentials inside the bulk region are indicated by dotted lines.

We first consider OBC. The resulting electrostatic potential [$V_{\text{OBC}}(y)$] is sketched by the stage-like curve (red line) in Figure 2. $V_{\text{OBC}}(y)$ can further be averaged inside the bulk region along the $y$ direction. Applying the "macroscopic average" as described by Baldereschi *et al.* [45], the average values at the positions $y_i$, $\bar{V}_{\text{OBC}}(y_i)$, are determined by integrating $V_{\text{OBC}}(y)$ over an interval of length $\alpha$ and dividing the result by $\alpha$:

$$\bar{V}_{\text{OBC}}(y_i) = \frac{1}{\alpha} \int_{y_i-\alpha/2}^{y_i+\alpha/2} V_{\text{OBC}}(y) \mathrm{d}y. \tag{8}$$

For the configurations with equidistant lattice planes, we chose $\alpha$ equal to twice the interplanar distance, i.e., the integration interval extends from $y_{i-1}$ to $y_{i+1}$. This corresponds to averaging out the microscopic fluctuations inside each unit cell on the order of a lattice parameter [46]. Our values $\bar{V}_{\text{OBC}}(y_i)$ follow a linear trend and can therefore be extrapolated by a line, which is indicated by the red dotted line in Figure 2. Such a linear average potential across the supercell with a non-zero slope corresponds to the presence of a surface dipole [33].

In the case of PBC, a vacuum of arbitrary size can be added to the set of atomic planes in the supercell, as sketched by the extension of the dashed blue box in Figure 1 in $y$ direction. The internal electric field imposed by the PBC (*cf.* Section 2) ensures the connection of the two end points of $V_{\text{OBC}}(y)$ extended into the vacuum region to the limits of the supercell (red curve). This transforms the stage-like curve into the zig-zag curve, as sketched by the green solid line in Figure 2. There is a resulting internal electric field inside the bulk region, if the slope of the macroscopic average potential (dotted green line) is non-zero. As in the case of OBC, it has its origin in the surface dipole. The magnitude of the internal electric field depends on the vacuum size. Two limiting cases can be considered: in the limit of an infinite vacuum length ($V_{\text{cell}} \to \infty$), the potential curve for PBC is identical to that of OBC. The other limit is a vacuum length of $d/2$ at both ends of the cell. Since this corresponds to no vacuum at all but a periodic bulk structure, there is no surface dipole, resulting in a flat average potential and therefore no internal electric field. In other words, the linear correction term for the potential in the case of PBC and a "vacuum" size of $d/2$ is identical to the average potential in the case of OBC.

### B. Charged supercell with equidistant, alternatingly charged planes

Next, we consider the scenario where one positively charged plane is added as the left termination plane to the stacking sequence of charged lattice planes treated in scenario A. In this case, the supercell is charged with $Q_{\text{tot}} = +q$.

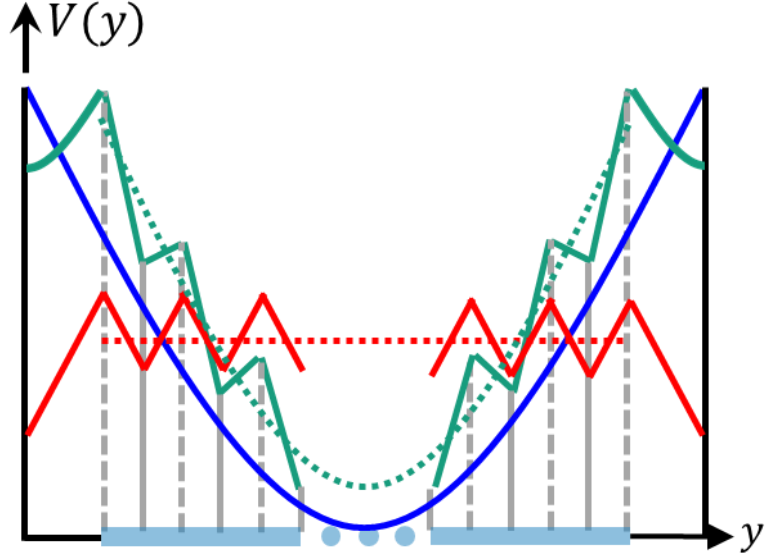

Figure 3. The electrostatic potential within a charged cell with equidistant, alternatingly charged planes [indicated by solid (negative) and dashed (positive) vertical gray lines] for OBC (red line) and PBC (green line). The corresponding macroscopic averages inside the bulk region are indicated by dotted lines. The blue parabola corresponds to the quadratic potential term originating from the neutralizing background charge density, with the minimum located at the center of charge.

Again, we first apply OBC, corresponding to a (single slab) supercell for the simulation of charged surfaces. Since the system is finite in $y$ direction, a neutralizing background charge density is not needed in this case ($\rho_0 = 0$). The resulting electrostatic potential is sketched by the zig-zag curve in Figure 3 (solid red line). In contrast to the situation in scenario A, the averaged potential is flat (dotted red line), because now there is no surface dipole between the two equally charged terminating surfaces.

In the case of PBC, corresponding to a periodically repeated slab system of infinite size, a neutralizing background charge ($-q$) needs to be included. Following Eq. (7), this leads to a quadratic term in the electrostatic potential as schematically sketched by the blue parabola across the supercell in Figure 3 with the minimum being at the center of charge, which coincides with the geometric center of the supercell in our setup. Combining the zig-zag potential [$V_{\text{OBC}}(y)$] with the quadratic term leads to a total electrostatic potential (shown in green), which is a parabola-like zig-zag curve in the bulk region and a parabola in the vacuum region. Using the averaged line of $V_{\text{OBC}}(y)$ inside the bulk, one obtains the averaged total potential (green dotted line).

C. <u>Charged cell with interfaces between two grains</u>

Finally, we consider a supercell with two crystal parts and an interface between them, e.g., a grain boundary (GB) between two differently oriented grains [Figure 4(a)]. In our example, both crystal parts, denoted by grain I (with $N_1$ lattice planes) and grain II (with $N_2$ lattice planes), are of the form described in scenario B, but with different distances ($d_1$, $d_2$) and different charges ($\pm q$, $\pm q'$) of the planes. The total charge of the supercell cell is then $Q_{\text{tot}} = q + q'$. The distance at the interface, denoted by $\delta_{\text{GB}}$, is typically a bit larger than the interplanar distances. As we describe in Section 4, such a supercell can be used for the simulation of an ATGB.

First, we consider OBC. Similar to the potential in scenario B, a zig-zag potential profile is obtained as sketched in Figure 4(b) (solid red curve). At the positions $y_i$ of each plane, the macroscopic average value $\overline{V}_{\text{OBC}}(y_i)$ was calculated following Equation (8). These averaged values follow a linear line in each grain (red dotted lines), which is not flat because of the presence of an interface dipole. To match the average potential lines of both grains in the interface region, they are extrapolated to $\tilde{y}_1 = y_1 - \frac{d_1}{2}$, $\tilde{y}_{N_1} = y_{N_1} + \frac{d_1}{2}$ and $\tilde{y}_{N_1+1} = y_{N_1+1} - \frac{d_2}{2}$, $\tilde{y}_{N_1+N_2} = y_{N_1+N_2} + \frac{d_2}{2}$, respectively, where they reach the not averaged stage-like potential curve. Here, we introduced a reference point $y_{\text{B}}^{\text{ref.}}$ somewhere in the bulk region of a grain, with the average potential $V_{\text{B}}^{\text{ref.}}$ at this point. As a result, the average potential within the supercell reads:

$$\overline{V}_{\text{OBC,d}}(y) = V_{\text{B}}^{\text{ref.}} + \frac{q'}{2A\epsilon_0}\left(y - y_{\text{B}}^{\text{ref.}}\right) \text{ for } \tilde{y}_1 \leq y \leq \tilde{y}_{N_1}, \tag{9-a}$$

$$\overline{V}_{\text{OBC,d}}(y) = V_{\text{B}}^{\text{ref.}} + \frac{q'-q}{2A\epsilon_0}\left(y - y_{\text{B}}^{\text{ref.}}\right) + \frac{q}{2A\epsilon_0}\left(\tilde{y}_{N_1} - y_{\text{B}}^{\text{ref.}}\right) \text{ for } \tilde{y}_{N_1} < y < \tilde{y}_{N_1+1}, \tag{9-b}$$

$$\overline{V}_{\text{OBC,d}}(y) = V_{\text{B}}^{\text{ref.}} - \frac{q}{2A\epsilon_0}\left(y - \tilde{y}_{N_1}\right) + \frac{q'}{2A\epsilon_0}\left(\tilde{y}_{N_1+1} - y_{\text{B}}^{\text{ref.}}\right) \text{ for } \tilde{y}_{N_1+1} \leq y \leq \tilde{y}_{N_1+N_2}. \tag{9-c}$$

The second subscript (d) in the potential indicates that the formalism is only valid for a system with equidistant planes in each of the bulk regions. Note, that the length of the interface separation ($\delta_{\text{GB}}$) does not influence the slopes of the average potential lines, but only the offset between them.

Next, we consider PBC. In this case the supercell contains two interfaces, or one interface and two surfaces, depending on the vacuum length. If we are only interested in the properties of one specific interface, as e.g., the interface energy, the two interfaces must be identical. This defines the supercell dimension [indicated by the blue dashed box in Figure 4(a)].

Following Equation (6), the total potential is obtained as:

$$V_{\text{PBC,d}}(y) = V_{\text{OBC,d}}(y) + \frac{q+q'}{2\epsilon_0 V_{\text{cell}}}y^2 - \sum_{i=1}^{N_1+N_2}\frac{q_i y_i}{\epsilon_0 V_{\text{cell}}}\cdot y + C. \tag{10}$$

The quadratic potential term results from the neutralizing background charge density $\rho_0 = -\frac{q+q'}{V}$ [dashed blue parabola in Figure 4(b)]. The linear potential term (dashed black line) is required by the periodicity, connecting the two ends of the stage-like red potential curve. Combining the linear and the quadratic potential terms leads to one parabola centered at the center of charge, which is vertically shifted with respect to the quadratic term in Eq. (10) [*cf.* Eq. (7)]. Adding the stage-like potential [$V_{\text{OBC,d}}(y)$] to it, we obtain the total potential $V_{\text{PBC,d}}(y)$ as sketched in Figure 4(c) by the solid green line. The dotted green line is its macroscopic average:

$$\overline{V}_{\text{PBC,d}}(y) = \overline{V}_{\text{OBC,d}}(y) + \frac{q+q'}{2\epsilon_0 V_{\text{cell}}}(y-\overline{y})^2 + C', \tag{11}$$

with $C' = C - \frac{q+q'}{2\epsilon_0 V_{\text{cell}}}\overline{y}^2$.

By applying scenario A to both grains instead of scenario B, we obtain a charge neutral supercell. An example would be a supercell of STO containing a symmetric tilt grain boundary [33]. Furthermore, the supercell of an interface can be constructed by combining one crystal of scenario A and another one of scenario B. An example is the heterojunction between neutral Ge and (001)-oriented polar GaAs [7].

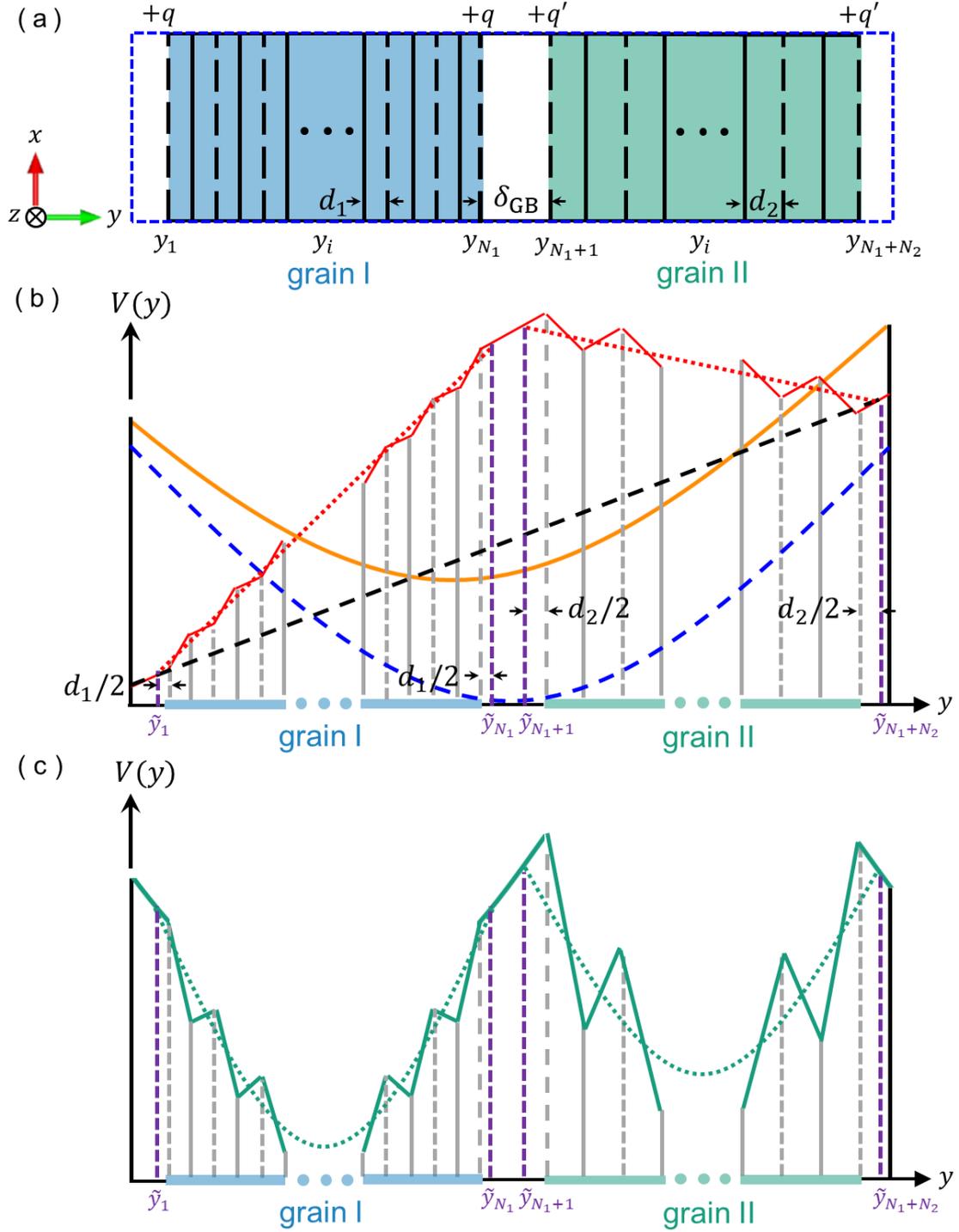

Figure 4. (a) A sketch of a supercell consisting of two grains (grain I, grain II), each with equidistant ($d_1$, $d_2$) and alternatingly charged ($\pm q$, $\pm q'$) planes. Both grains have equally charged termination planes, and are separated by $\delta_{\mathrm{GB}}$ at the interface; (b) the resulting electrostatic potential for OBC (red curve, with macroscopic average indicated by the dotted line), together with the linear potential introduced when applying PBC (dashed black line), the quadratic potential originating from the background charge (dashed blue curve), and the shifted parabola (solid orange curve) from combining the latter two terms [*cf.* Equation (7)]; (c) the resulting electrostatic potential for PBC (solid line) and its macroscopic average (dotted line).

## 4   Application of the model to atomistic calculations

For a system described by an empirical interatomic potential, we demonstrate in this section, how the electrostatic potential within a supercell consisting of charged planes affects the formation energies of charged point defects and the relaxed structure. Specifically, we consider positively charged oxygen vacancies in the cubic perovskite STO with an ATGB (430)||(100). For the force field, we apply a rigid-ion model with parameters optimized by Thomas *et al.* [47]. This "Thomas potential" has be shown to reasonably describe grain boundaries in STO in atomistic simulation (see Refs. [33,48] and references therein). In the model, the ions are given partial charges deviating from the formal oxidation states. As will be described in more detail in Section 4.1, this leads to charged planes corresponding to Tasker's type 3 and to a totally charged supercell of the considered system.

The resulting internal potential could in principle be "measured" by moving a probe charge through the cell, which only interacts electrostatically, and simultaneously calculating the total cell energy. A charged oxygen vacancy serves as such a test charge in this study. Since the total cell energy itself is quantitatively not meaningful, we reference it to the energy of a cell in which the vacancy is located deep in the bulk of one of the grains, i.e., far from the interfaces. In this way, we obtain the difference in formation energies of an oxygen vacancy located somewhere in the cell, for instance at the GB, and one located in the bulk. This energy difference can then be analyzed, e.g., with respect to the GB type or structure, or to the composition of the material. However, for a correct interpretation, the artificial field encountered by the vacancy must be subtracted, since it is not present in a realistic material. Hence, the expression for the corrected formation energy ($\Delta E_\text{f}$) in this model as a function of the position $y$ in the direction perpendicular to the GB becomes:

$$\Delta E_\text{f}(y) = E_\text{tot}(y) - E_\text{tot}\big(y_\text{B}^\text{ref.}\big) - q^{V_\text{O}}\big[\overline{V}(y) - V_\text{B}^\text{ref.}\big]. \tag{12}$$

$E_\text{tot}$ denotes the total cell energy, $q^{V_\text{O}}$ the charge of the oxygen vacancy, $\overline{V}$ the averaged electrostatic potential as, e.g., given in Eqs. (9) and (11), $y_\text{B}^\text{ref.}$ is the reference point in the bulk and $V_\text{B}^\text{ref.} = \overline{V}\big(y_\text{B}^\text{ref.}\big)$.

Energy and force calculations are performed with the program GULP [General Utility Lattice Program (GULP)]. The procedure is first applied to an unrelaxed structure for both, OBC and PBC, and then to a relaxed structure for PBC, followed by a discussion of the effect of the artificial electric field on the structure during relaxation.

### 4.1   The unrelaxed ATGB structure

We constructed a supercell of STO containing an ATGB with the orientation relationship (430)[001]||(100)[001], as shown in Figure 5. As in the generic scenarios described in Section 3, we chose the $y$ axis (lattice parameter $b_\text{cell}$) perpendicular to the interface, here, the GB plane. The periodic length in the (430) plane is five times of that in the (100) plane, hence the cell parameter in $x$ direction ($a_\text{cell}$) is set to $5a_\text{STO}$ (which is the length of the primitive cell of the coincidence site lattice, $a_\text{CSL}$), with $a_\text{STO}$ denoting the lattice constant of STO ($a_\text{STO} = 3.905\,\text{Å}$ for the Thomas potential). We set the cell parameter in $z$ direction, $c_\text{cell}$, to $1a_\text{STO}$. With those

choices, the $x$-$z$ planes have an area $A$ of 76.25 Å$^2$. The length of the (430) oriented grain in $y$ direction is set $20a_{STO}$, corresponding to $4a_{CSL}$, and the length of the (100) oriented grain is set to $12a_{STO}$. The two grains are put together with a separation of $\delta_{GB} = a_{STO}/2$. A "vacuum" of length $a_{STO}/4$ is added at both ends of the supercell (with respect to $y$) to ensure a reasonable structure if applying PBC, which results in a total cell parameter $b_{cell}$ of $33a_{STO}$. The size is considered large enough to clearly distinguish the GB and surface regions from the bulk parts of the grains in the further analysis, e.g. of defect formation energies. An optimization of the GB structure by performing rigid body translations of the grains with respect to each other was not performed, since this would not change the plane orientations and is therefore not relevant for the purpose of this study.

Each grain consists of alternating $x$-$z$ planes: a Sr-O and a Ti-O$_2$ plane in the (430) oriented grain, and a Sr$_5$-O$_5$ and a Ti$_5$-O$_{10}$ plane in the (100) oriented grain. In the Thomas potential, the Sr, Ti, and O ions are given the effective charges $+1.84e$, $+2.36e$ and $-1.4e$, respectively, with e denoting the elementary charge, which leads to total plane charges of $+0.44e$ (Sr-O), $-0.44e$ (Ti-O$_2$), $+2.2e$ (Sr$_5$-O$_5$), and $-2.2e$ (Ti$_5$-O$_{10}$). To ensure two equivalent GBs if applying PBC, the two grains are each terminated by the same (positively charged) plane at both ends. Altogether, the structure is a specific example of the generic case described in scenario C, with "grain I" being the (430) and "grain II" the (100) oriented grain, and, correspondingly, with the interplanar distances $d_1 = 0.3905$ Å and $d_2 = 1.9525$ Å. The cell consists of 166 Sr, 160 Ti, and 486 O ions, leading to a total charge $Q_{tot}$ of $+2.64e$.

Note that the individual planes in both grains, and hence the total cell, would be charge neutral if formal oxidation states were considered instead of the partial charges of the Thomas potential, in which case no electrostatic potential would be present in the cell (Tasker's type 1). However, if considering cubic perovskites, this only holds for II-IV compounds (such as Sr$^{+II}$Ti$^{+IV}$O$_3$). In contrast, III-III or I-V systems, such as La$^{+III}$Al$^{+III}$O$_3$ or K$^{+I}$Ta$^{+V}$O$_3$, possess charged planes of the given orientations for the formal ionic charges, too.

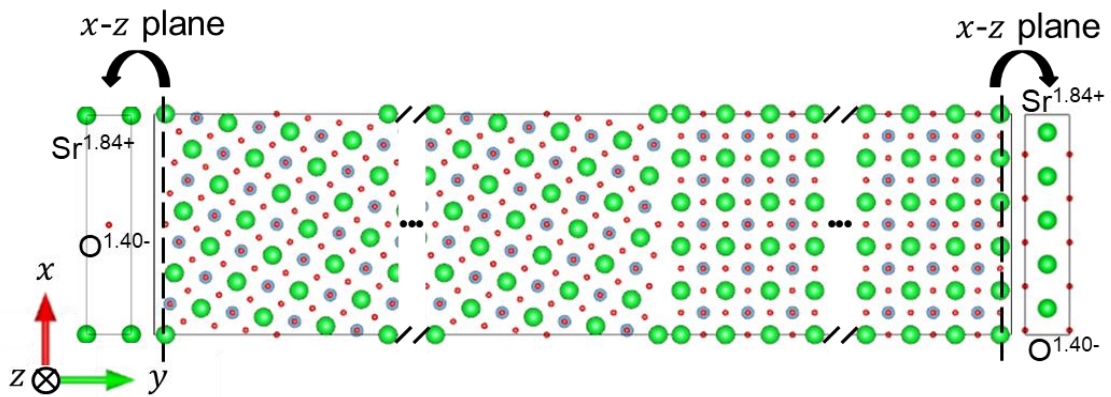

Figure 5. The unrelaxed structure of a supercell of STO containing an ATGB (430)||(100), viewed from the [001] tilt-axis direction. Sr ions are displayed by large green, Ti atoms by intermediate blue, and oxygen ions by small red spheres. The $x$-$z$ planes are alternatingly positively or negatively charged. The ionic compositions leading to those charges are exemplarily shown for the surface planes at the two ends of the supercells.

In both grains, several repeated bulk units are omitted for the sake of clarity, as indicated by the black dots.

Oxygen ions were removed from the STO supercell from all possible oxygen sites one-by-one, and the respective vacancy formation energies were calculated as described in the beginning of Section 4. First, OBC were applied. The reference point $y_\text{B}^\text{ref.}$ was chosen in the center of the bulk region of the (430) oriented grain. The values obtained without applying the electrostatic correction are displayed in Figure 6(a) (labeled as "simulation data"). The effect of the electrostatic potential is clearly visible by a strong linear variation of the values around the reference point by about $\pm 150$ eV across the (430) bulk grain, and a weaker linear, but still considerable variation by about $\pm 18$ eV around the midpoint across the (100) bulk grain.

Inserting the parameters reported above in this subsection, the electrostatic potential $\overline{V}_\text{OBC,d}(y)$ was determined according to Eqs. (9-a)–(9-c). With $q^{V_\text{O}} = +1.40e$ and $\overline{V}(y) = \overline{V}_\text{OBC,d}(y)$, the correction part of Eq.(12) was calculated, which is shown in Figure 6(a), labelled as "model function". In the bulk regions, the simulation data points deviate from this function on average only on the order of $0.01$ eV, which confirms the validity of the model. The corrected datapoints clearly show the influence of the GB on the oxygen vacancy formation energy.

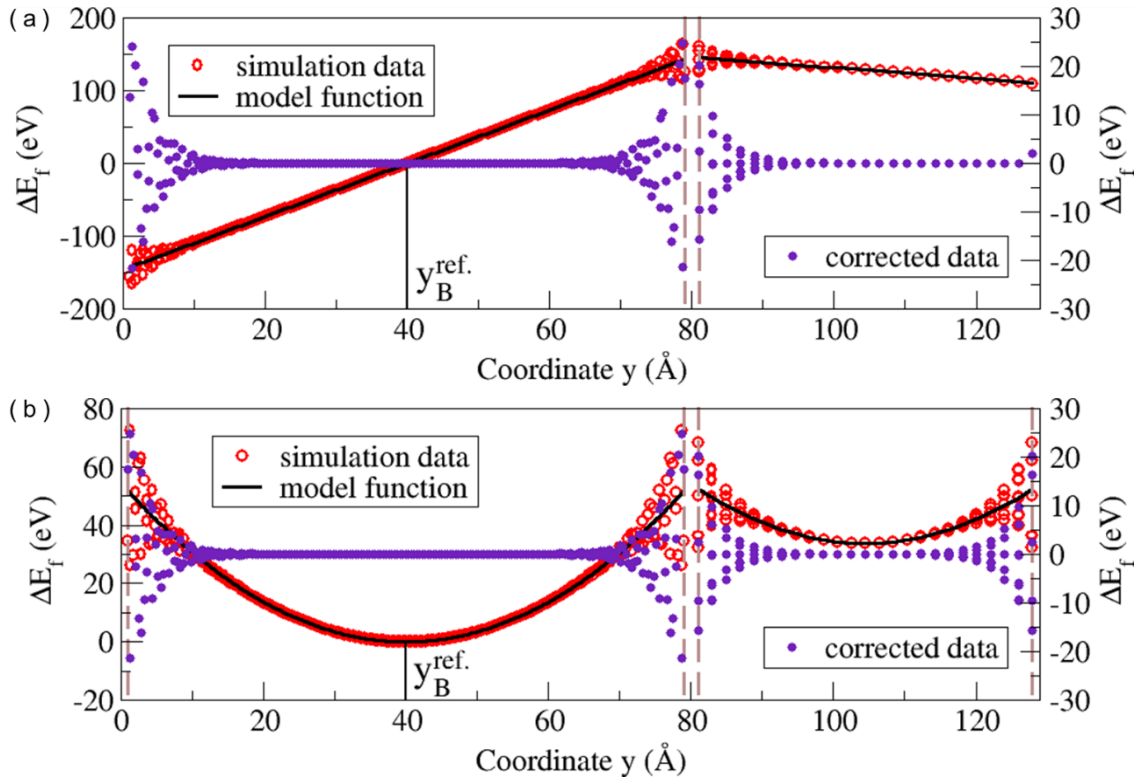

Figure 6. The formation energies of positively charged oxygen vacancies ($\Delta E_\text{f}$) in the unrelaxed supercell of STO containing an ATGB (430)||(100) for (a) OBC and (b) PBC. The simulation data (left energy axis) match the line of the analytical model function. The values corrected for the internal electrostatic potential are plotted with respect to the rescaled right energy axis for a better visibility. The dashed vertical lines indicate the GB regions.

Next, PBC were applied. Since the cell is charged, the "simulation data" in Figure 6(b) show a parabolic behavior in both bulk regions. Using as the potential $\overline{V}(y)$ now the expression for $\overline{V}_{\text{PBC,d}}(y)$ from Eq. (11) with $\overline{V}_{\text{OBC,d}}(y)$ from Eqs. (9-a)–(9-c), the model function, and, with it, the corrected formation energies, were determined from Eq. (12) in the same way as described for OBC. Clearly, the model reproduces the internal potential encountered by the charged defect for PBC, too.

## 4.2 The relaxed ATGB structure

### 4.2.1 Derivation of the potential function

Relaxation of the atomic positions in the structure described in Section 4.1 was performed at constant cell volume for both, OBC and PBC. For OBC, the grains were found to preserve their bulk structure showing only slight ionic displacements close to the surfaces. Therefore, the oxygen vacancy formation energies showed largely the same behavior as discussed for the unrelaxed structure. In the following, we will focus on the relaxed structure obtained for PBC. The structure around the ATGB is displayed in Figure 7.

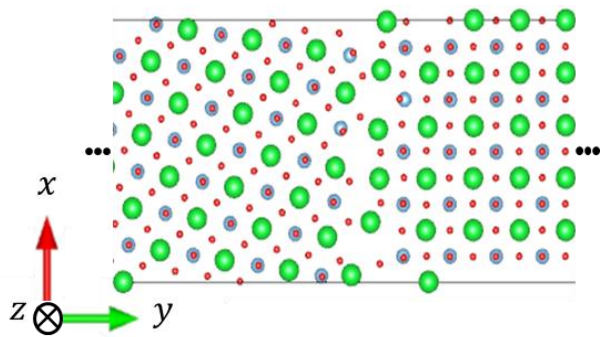

Figure 7. The structure of the ATGB(430)||(100) in STO, viewed from the [001] direction, after structural relaxation with PBC. Compared to Figure 5, only the central part of the supercell is displayed. Sr ions are colored green, Ti ions blue and oxygen ions red.

In the relaxation, the ions were displaced from their perfect positions on the $x$-$z$ planes. While in the bulk regions of the two grains, the deviations mainly occurred in $y$ direction by less than $0.1$ Å, some ions close to the GBs were displaced by up to $0.2$ Å.

To probe the electrostatic potential in this structure, we again calculated the formation energies of oxygen vacancies on all possible sites. The uncorrected "simulation data" shows a parabolic behavior in both grains, as displayed in Figure 8. However, the parabolas are not equivalent to those obtained for the unrelaxed structure [*cf.* Figure 6(b)] because of the displaced (relaxed) ions. Instead of well-defined $x$-$z$ planes comprising ions in the stoichiometric compositions described in Section 4.1, now, each ion has a different $y$ coordinate and therefore needs to be considered as a separate, charged $x$-$z$ plane for the potential summation. Equation (7) is generally valid, i.e., it still describes the total potential $V_{\text{PBC}}(y)$ and, accordingly, the plane-averaged $\overline{V}_{\text{PBC}}(y)$. But since the planes are not equidistant anymore, the term $\overline{V}_{\text{OBC}}(y)$ entering the expression for $\overline{V}_{\text{PBC}}(y)$ no longer follows Equations (9-a)–(9-c).

To determine an expression for $\overline{V}_{\text{OBC}}(y)$ for the relaxed structure, we first determined the accurate potential function $V_{\text{OBC}}(y)$ across the supercell according to Equation (5). At the positions $y_i$ of the planes, the macroscopic average values $\overline{V}_{\text{OBC}}(y_i)$ were then calculated following Equation (8), with integration intervals $\alpha_i$ extending from $y_i - d_{i,\min}$ to $y_i + d_{i,\min}$, where $d_{i,\min}$ denotes the shorter one of the two distances between $y_i$ and its neighboring planes. In contrast to the linear behavior for the unrelaxed configuration of equidistant planes, in each of the two bulk regions, the values $\overline{V}_{\text{OBC}}(y_i)$ obtained in this way showed a quadratic behavior. We finally obtained the functions $\overline{V}_{\text{OBC}}(y)$ by fitting those values to quadratic polynomials, separately for each grain. A qualitative explanation for the quadratic behavior is given in Section 4.2.2.

Inserting the functions $\overline{V}_{\text{OBC}}(y)$ into the plane-averaged version of Eq. (7) and using $\overline{V}_{\text{PBC}}(y)$ derived in this way in Eq. (12), we obtain the model function of the internal electrostatic potential, and, with it, the corrected vacancy formation energies, as shown in Figure 8.

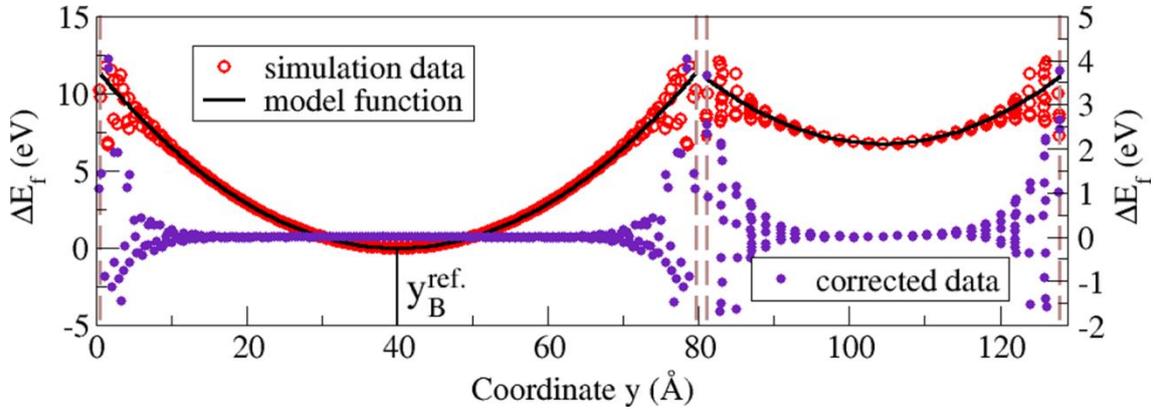

Figure 8. The relative formation energies of oxygen vacancies ($\Delta E_f$) in a supercell of STO containing an ATGB (430)||(100), which was relaxed with PBC, before ("simulation data", left axis), and after ("corrected data", right axis) applying the electrostatic correction with the analytical "model function" (black line). The dashed vertical lines indicate the GB regions.

### 4.2.2 Influence of the potential on relaxation

The origin of the parabolic behavior of $\overline{V}_{\text{OBC}}(y)$ for the structure relaxed with PBC as described in the previous subsection can be understood by considering the repeat units of oppositely charged planes in each grain. As described in Section 4.1, before the relaxation, the (430) oriented grain consists of equidistant, alternating positively charged Sr-O ($+0.44\,e$) and negatively charged Ti-$O_2$ planes ($-0.44\,e$). Upon structural relaxation, the ions experience displacements. In the GB regions, they are dominated by the forces between the ions to optimize the GB structure. In the bulk regions of the grains, where the influence of the GB is negligible, the displacements mainly result from the interaction of the ions with the artificial internal electrostatic potential. For this case, two displacement mechanisms can be distinguished: First, the differently charged ions within each of the Sr-O and Ti-$O_2$ planes are displaced from these planes, however, by less than $0.1$ Å. Hence, we treat the relaxed pair of one Sr and one O ion, and the relaxed triple of one Ti and two O ions each as one ionic group, and we define the position of such a group as the mean value of the ions belonging to it. Second, the averaged positions of these groups, specifically, their relative distances to the neighboring groups, are affected by the electrostatic potential, which is the reason behind the parabolic behavior of $\overline{V}_{\text{OBC}}(y)$. Let the distance between the Sr-O group $(+q)$ and the neighboring Ti-O$_2$ group $(-q)$ in positive $y$ direction be $D_1$, and the distance between a Ti-O$_2$ group and the Sr-O group in the next repeated ionic group in the same direction be $D_2$.

In the initial structure with equidistant planes, the parabolic nature of the potential, solely originating from the homogeneous background charge, has its minimum in the center of the cell, i.e., the electric field is zero at that point [*cf.* Figure 4(c)]. The further away a repeat unit of two oppositely charged planes (or ionic groups in the later steps of the relaxation) in the sequence $(+q,-q)$ is from the midpoint of the grain, the larger is the magnitude of the electric field acting on the planes, resulting in a stronger movement of the planes in opposite directions, approaching each other: $D_1$ decreases and $D_2$, the distance to the next repeat unit, increases, as exemplarily illustrated in Figure 9(a). This builds up a depolarization field counteracting the internal electric field. However, the latter cannot be fully compensated by the induced depolarization field due to additionally active repulsive forces between the ions. Such rearranged ionic groups further adapt the profile of the internal potential [according to Eqs. (7) and (5)], which adjusts the positions of the repeated ionic groups in turn, until the structure reaches an energy minimum.

Since the electric field is linear, the distances $D_1$ and $D_2$ are both affected linearly as a function of the position during the relaxation. However, the sum $D = D_1 + D_2$ stays approximately constant ($\sim 2d$) because the structural relaxation was performed at constant volume. Figure 9(b) exemplarily shows the profile of the electrostatic potential, $V_{\text{OBC}}(y)$, for a sequence of planes stacked in the described way [Figure 9(a)]. Since the potential increment linearly decreases within each constant distance $D$ with respect to the $y$ coordinate, this results in a plane-averaged parabolic behavior of the potential. This motivates the choice of the quadratic fit functions for obtaining expressions for $\overline{V}_{\text{OBC}}(y)$ in Section 4.2.1.

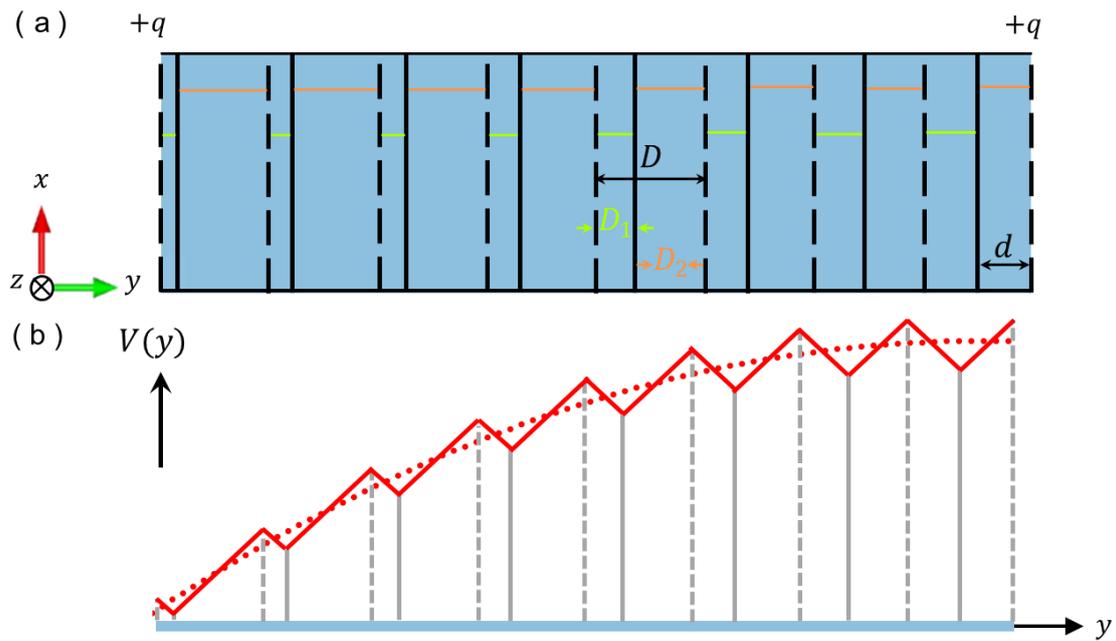

Figure 9. (a) A qualitative sketch of the stacking sequence as the one observed in the relaxed ATGB structure, consisting of unequally spaced, alternatingly charged ionic groups [indicated by dashed ($+q$) and solid ($-q$) vertical line] with linearly changing relative distances $D_1$ and $D_2$ (shown in light green and orange, respectively) and a constant sum $D = D_1 + D_2$ as a function of $y$. The right end of the stacking sequence corresponds to the center of a grain (bulk region), where the potential is minimal and $D_1 \approx D_2 \approx d$, the distance of the planes in the initial, unrelaxed structure. (b) The qualitatively resulting electrostatic potential $V_{\text{OBC}}(y)$ (solid red line), and the corresponding plane-averaged function (dotted red line).

To avoid the artificial ionic displacements occurring during the relaxation due to the existence of the internal electrostatic potential, the forces acting on the ions need to be corrected at each simulation step by subtracting the contribution stemming from the electric field. If this is not done, as in the examples described here, the final structure must be carefully evaluated based on the provided analysis in order to avoid interpretations which are physically not meaningful. But as long as, for an unoptimized structure, the artificial displacements take place on a much smaller length scale compared to those originating from the interatomic potential, major structural features such as the energetically most stable ionic configuration at a GB, are rather unaffected by the effect of the uncompensated electrostatic potential.

## 5    Summary and conclusions

We investigated the origin and analyzed the effect of electrostatic potentials which are present in finite atomistic supercells of ionic crystals with interfaces, which are described by classical interatomic models. The potentials are caused by the stoichiometric, charged interface planes and by the total charge of the supercell. Since they are compensated in real materials, the existence of these potentials in the simulation model is artificial and needs to be correctly treated in the analysis. To this end, we analytically derived one-dimensional expressions for the potentials for both, open and periodic boundary conditions, and systematically applied them to three generic examples containing alternatingly charged lattice planes with increasing complexity: a neutral cell with equidistant planes, a charged cell with equal termination planes, and a cell consisting of two different and separated grains. The application of open boundary conditions led to linear potentials across the supercells for each scenario, corresponding to the existence of surface and grain boundary dipoles. With periodic boundary conditions, the same effects are obtained for neutral cells, but in the case of charged cells, a parabolic potential is additionally present arising from the compensating background charge. Since this corresponds to a linearly changing electric field, the ions in the supercell experience artificial, non-constant electrostatic forces and displace accordingly upon relaxation. We validated our formalism by applying it to supercells of the cubic perovskite strontium titanite with an asymmetric tilt grain boundary. The effect of the artificial electrostatic potential across the supercell in the direction perpendicular to the grain boundary was probed by positively charged oxygen vacancies, and their formation energies were corrected using our derived potential functions.

The developed model can be applied to other charged extended defects in ionic crystals, such as the heterojunctions treated in Refs. [7] and [45]. The correction method for the defect calculations can naturally be extended from oxygen vacancies to cationic defects, e.g. strontium vacancies which were reported to be predominant in STO [49]. Even though developed and demonstrated for systems of ions interacting by classical rigid-ion-model potentials, the gained insight into the origin of electrostatic potentials in supercells with different boundary conditions may be relevant for strongly ionically bonded compounds treated by density functional theory as well. Altogether, the analytical correction method presented in this paper extends the family of general correction schemes for charged defects developed so far [50], towards the simulation of both charged point defects and charged extended defects inside polycrystals.

**Acknowledgements**

This work was funded by the German Research Foundation (DFG); Grants No. MR22/6-1 and EL155/31-1 within the priority programme "Fields Matter" (SPP 1959). Computations were carried out on the bwUniCluster computer system of the Steinbuch Centre of Computing (SCC) of the Karlsruhe Institute of Technology (KIT), funded by the Ministry of Science, Research, and Arts Baden Württemberg, Germany, and by the DFG. Structure figures were prepared with VESTA [51].